\documentclass[hidelinks,twocolumn, prl, superscriptaddress,a4paper]{revtex4}

\usepackage[colorlinks=true,allcolors=blue]{hyperref}

\usepackage{graphicx}
\usepackage{dcolumn}
\usepackage{orcidlink}
\usepackage{csquotes}
\usepackage{float}
\usepackage{amsmath}
\usepackage{braket}
\usepackage{bm}
\begin{document}

\preprint{APS123-QED}


\title{Spread Complexity in Non-Hermitian Many-Body Localization Transition}


\author{Maitri Ganguli\,\orcidlink{0009-0009-4701-2459}}
\email[]{maitrig@iisc.ac.in}
\affiliation{Department of Physics, Indian Institute of Science, Bangalore}


\date{\today}

\begin{abstract}
We study the behavior of spread complexity in the context of non-Hermitian many-body localization transition (MBLT). Our analysis has shown that the singular value spread complexity is capable of distinguishing the ergodic and many-body localization (MBL) phase from the presaturation peak height for the non-hermitian models having time-reversal symmetry (TRS) and without TRS. On the other hand, the saturation value of the thermofield double (TFD) state complexity can detect the real-complex transition of the eigenvalues on increasing disorder strength. From the saturation value, we also distinguish the model with TRS and without TRS. The charge density wave complexity shows lower saturation values in the MBL phase for the model with TRS. However, the model without TRS shows a completely different behavior, which is also captivated by our analysis. So, our investigation unravels the real-complex transition in the eigenvalues, the difference between the model having TRS and without TRS, and the effect of boundary conditions for the non-hermitian models having MBL transitions, from the Krylov spread complexity perspective.
\end{abstract}


\maketitle

\section{Introduction}
The study of quantum complexity in the Krylov space has gained a lot of attention in recent days, in the context of both states and operators, see \cite{Nandy:2024htc} for a review. The idea of Krylov complexity has been applied to a wide range of topics, from operator growth, spread complexity in spin chains, SYK model \cite{Parker_2019,Camargo:2024deu,Balasubramanian:2022tpr,Rabinovici_2021,Bhattacharjee_2023,Chapman:2024pdw,Bhattacharjee_2022,Bhattacharya:2023xjx,Bhattacharya:2022gbz,Bhattacharjee_2024,Bhattacharya:2023zqt,Bhattacharya:2023yec,Bhattacharya:2024hto} to random unitary circuits \cite{suchsland2023krylovcomplexitytrottertransitions,Sahu:2024urf}, QFTs \cite{Avdoshkin_2024,Camargo_2023,Vasli_2024,he2024probingkrylovcomplexityscalar}, CFTs \cite{PhysRevD.104.L081702,Kundu_2023} and string scattering amplitudes \cite{Bhattacharya:2024szw} etc. The key motivation in all of these investigations is to quantify chaos from the perspective of Krylov complexity. In the context of operator growth, the Universal operator growth hypothesis~\cite{Parker_2019} states that the Lanczos coefficients grow as fast as possible (linear) in chaotic systems. Moreover, the Krylov exponent (rate of exponential growth of Krylov complexity) acts as a bound on the Lyapunov exponent calculated from OTOCs and generalizes the Maldacena-Shenker-Stanford (MSS) bound~\cite{Maldacena_2016}. On the other hand, the role of the spread complexity of states in distinguishing chaotic from integrable has been investigated recently. In particular, the presence of a peak before saturation for Thermofield Double (TFD) state spread complexity has been established as a signature of chaos, which is verified in different Random matrix theory (RMT) ensembles as well as various quantum many-body systems~\cite{Balasubramanian:2022tpr,Camargo:2024deu}. Additionally, the peak height can be used as an order parameter in the chaotic to integrable phase transitions, for example, in mass-deformed SYK model~\cite{Baggioli:2024wbz} and in Many-body localization (MBL) transition~\cite{Ganguli:2024myj}.

In interacting quantum many-body systems the presence of strong disorder can give rise to localized energy eigenstates, emergent integrability due to an extensive number of quasi-local integrals of motion (LIOMs) and as a consequence violation of eigenstate thermalization hypothesis (ETH)~\cite{Srednicki_1994,Pal_2010,Nandkishore_2015}. This phase is known as the Many-body localized (MBL) phase~\cite{Abanin_2019}. Krylov operator complexity in the MBL phase has already been studied in Ref.~\cite{Trigueros:2021rwj}. In a recent study in Ref.~\cite{Ganguli:2024myj}, initial state-dependent spread complexity has been analyzed in the context of MBL transition for both closed and open quantum systems. It has been found that the peak height in the TFD complexity (which can be treated as an order parameter) goes to zero during the MBL transition and the the saturation value of an initial ordered state spread complexity is also indicative of the appearance of LIOMs in the MBL phase. In the present work, our goal is to extend these ideas in the context of non-Hermitian models which also exhibit MBL transition~\cite{Hamazaki_2019,PhysRevB.101.184201,Li_2023,Suthar_2022}.

Delocalization to localization transition driven by increasing disorder strength also occurs in systems with non-unitary evolution (generated by underlying non-Hermitian Hamiltonian), which is known as the non-Hermitian MBL transition. Such non-unitary dynamics are commonplace in the open quantum system dynamics (e.g, current or voltage driven or continuously measured systems) of a single quantum trajectory with no quantum jumps~\cite{doi:10.1080/00018732.2014.933502}, pure states with post-selected measurement outcomes. MBL transition in non-Hermitian interacting disordered systems can be understood from the transition of entanglement entropy as well as entanglement dynamics, see for example~\cite{Hamazaki_2019,PhysRevB.101.184201}. In addition, if the non-Hermitian model possesses time-reversal symmetry (TRS), then a complex (eigenvalues) to real transition also happens by increasing disorder strength, which may not coincide with the MBL transition point~\cite{PhysRevB.101.184201}. On the other hand, if the model is not time-reversal invariant, then there won't be any complex-real transition, still a many-body localization transition can occur~\cite{Hamazaki_2019}.

In this work, our motivation is to understand non-Hermitian MBL phenomena as well as complex-real transitions from the perspective of spread complexity dynamics. We have considered disordered interacting hard-core boson models with asymmetric hopping (with TRS) and onsite particle loss-gain (without TRS) with both periodic boundary conditions (PBC) and open boundary conditions (OBC). To understand the chaotic-to-integrable transition from the peak height of spread complexity, we introduce the notion of singular value spread complexity which is inspired by the notions of singular value form factor and singular value spectral complexity introduced in Ref.~\cite{nandy2024probingquantumchaossingularvalue}. Our results show that the peak height in the singular value spread complexity can be used as an order parameter for chaotic-to-integrable transitions in non-Hermitian systems. We also study the spread complexity of TFD states (for which we have to use Arnoldi iteration instead of Lanczos algorithm since the Hamiltonian is non-Hermitian), and the results indicate that the presence of complex energy eigenvalues cause the TFD spread complexity to saturate below \(0.5\). That is, if the eigenvalues are all real, then only the saturation reaches \(0.5\). Therefore, the saturation value in TFD spread complexity is indicative of complex-real transition. To understand the localization transition, we take the Charge density wave (CDW) state as the initial ordered state and compute its complexity dynamics. In the model with TRS, a lesser saturation value without any peak is observed in the MBL phase, in contrast to the presence of a peak/higher saturation value in the ergodic phase. In the model without TRS, the results are qualitatively different from the case with TRS. In this case, the CDW spread complexity saturation is higher without any peak (or with a small peak) in the MBL phase and in the ergodic phase, the complexity saturation value is lesser with a sharp pre-saturation peak.

The rest of the paper is organized as follows. In Sec.~\ref{sec: models} we describe the models that we have worked with. In Sec.~\ref{sec: formalism} we introduce the notion of singular value spread complexity and describe the formalism used to calculate the singular value spread complexity, TFD complexity and CDW spread complexity. In Sec.~\ref{sec: results} we describe the results obtained for the different kinds of complexities introduced in the earlier section and how they can distinguish different quantum phases in non-Hermitian systems. In Sec.~\ref{sec: discussion} we conclude with the significance of our results and its future implications.

\section{The Models}\label{sec: models}

In this work, we have focused on two interacting non-Hermitian models that exhibit many-body localization transitions, one with time-reversal symmetry (TRS) and another without TRS~\cite{Hamazaki_2019}.

The model with TRS is a one-dimensional disordered interacting hard-core boson model with asymmetric hopping,

\begin{equation}
    \hat{H} = \sum_j \left[ -J (e^{-g} b_{j+1}^\dagger b_j + e^g b_{j}^\dagger b_{j+1} ) + U n_{j+1} n_j + h_j n_j \right]
\end{equation}

here \(b_i\) is the annihilation operator for hard-core boson, and \(n_i = b_i^\dagger b_i\) is the number operator at site \(i\). \(J\) is hopping parameter and the parameter \(g\) controls the assymmetry in the left-to-right and right-to-left hopping. \(U\) is the density-density interaction strength and \(h_i\) are random numbers sampled uniformly from the interval \([-h,h]\). We shall work in the half-filling sector with both periodic boundary conditions (PBC) and open boundary conditions (OBC).

To have a model without TRS, we consider again a one-dimensional disordered interacting hard-core boson model, but with onsite particle gain and loss,

\begin{equation}
\small
    \hat{H} = \sum_j \left[ -J (b_{j+1}^\dagger b_j + h.c.) + U n_{j+1} n_j + (h_j - i\gamma (-1)^j) n_j \right]
\end{equation}

in this model \(\gamma\) controls the non-Hermiticity. In this model, we also consider both PBC and OBC.

In our analysis, we have worked in the half-filling sector and with \(J=1,U=2,g=0.1,\text{ and }\gamma=0.1\). Only the disorder strength \(h\) will be varied.

\section{Spread Complexity in non-Hermitian Systems}\label{sec: formalism}

The Spread complexity of TFD states has been proven to be a useful measure to distinguish chaotic from integrable transitions. In this work, we seek generalizations of TFD spread complexity in non-Hermitian systems. We have used two generalizations, one using the singular values of the non-Hermitian Hamiltonian, which we call Singular Value Spread Complexity, and another using the actual complex eigenvalues, as a direct generalization of Hermitian models, which we call the usual TFD spread complexity. We also consider the spread complexity of an initial charge density wave (CDW) state evolving under non-Hermitian Hamiltonian, to probe localization properties of non-Hermitian MBL systems.

\subsection{A. Singular Value Spread Complexity}
We first describe the notion of singular value spread complexity, which utilizes the singular values of a non-Hermitian Hamiltonian. Suppose, for a \(d\)-dimensional system, the singular values are, \(\sigma_1,\sigma_2,\dots,\sigma_d\), which are real and non-negative by definition. Then we consider a fictitious Hermitian Hamiltonian, we call it the singular Hamiltonian, which we denote by \(\hat{H}_\sigma\), whose eigenvalues are the singular values of the actual non-Hermitian Hamiltonian. Denote the eigenvectors of \(\hat{H}_\sigma\) by \(\ket{\alpha}_\sigma\) with eigenvalues \(\sigma_\alpha\) where \(\alpha=1,2,\dots,d\). That is,
\begin{equation}
    \hat{H}_\sigma \ket{\alpha}_\sigma = \sigma_\alpha \ket{\alpha}_\sigma
\end{equation}
Then we compute the spread complexity of the following initial state,
\begin{equation}
    \ket{\psi}_\sigma = \left(1/\sqrt{d}\right) \sum_{\alpha=1}^d \ket{\alpha}_\sigma
\end{equation}
under the time-evolution with \(\hat{H}_\sigma\). The time-evolved state is,
\begin{equation}
    \ket{\psi (s)}_\sigma = e^{-i \hat{H}_\sigma s} \ket{\psi}_\sigma = \left(1/\sqrt{d}\right) \sum_{\alpha=1}^d e^{-i \sigma_\alpha s} \ket{\alpha}_\sigma
\end{equation}
where \(s\) is the singular time.

We are interested in calculating the Krylov spread complexity of the above singular time evolution. The first step is to construct the orthonormal singular Krylov basis vectors, \(\{\ket{K_n}_\sigma;\,n=0,1,2,\dots\}\) which span the Krylov space generated by \(\{\ket{\psi}_\sigma,\hat{H}_\sigma\ket{\psi}_\sigma,\hat{H}_\sigma^2\ket{\psi}_\sigma,\dots\}\). This can be achieved by applying the Lanczos algorithm that tridiagonalizes the singular Hamiltonian \(\hat{H}_\sigma\). We describe the algorithm below.

Define the first basis vector \(\ket{K_0}_\sigma\) to be \(\ket{\psi}_\sigma\) and the first Lanczos coefficient \(a_0\) by \(a_0 = {}_\sigma\bra{K_0}\hat{H}_\sigma\ket{K_0}_\sigma\). Then we do the following to evaluate \(\ket{K_1}_\sigma\) and to define \(a_1\), \(b_1\).
\begin{equation}
\begin{aligned}
    &\ket{A_1}_\sigma = \hat{H}_\sigma \ket{K_0}_\sigma - a_0 \ket{K_0}_\sigma\\
    &b_1^2 = {}_\sigma\bra{A_1}A_1\rangle_\sigma\\
    & \ket{K_1}_\sigma = \frac{1}{b_1} \ket{A_1}_\sigma\\
    & a_1 = {}_\sigma\bra{K_1}\hat{H}_\sigma\ket{K_1}_\sigma
\end{aligned}
\end{equation}

To get the other Krylov basis vectors and the Lanczos coefficients, we do the following for \(n>1\),
\begin{equation}
\begin{aligned}
    &\ket{A_n}_\sigma = \hat{H}_\sigma \ket{K_{n-1}}_\sigma - a_{n-1} \ket{K_{n-1}}_\sigma - b_{n-1}\ket{K_{n-2}}_\sigma\\
    &b_n^2 = {}_\sigma\bra{A_n}A_n\rangle_\sigma\\
    & \ket{K_n}_\sigma = \frac{1}{b_n} \ket{A_n}_\sigma\\
    & a_n = {}_\sigma\bra{K_n}\hat{H}_\sigma\ket{K_n}_\sigma
\end{aligned}
\end{equation}

Once the Krylov basis vectors are constructed, we can expand the time-evolved state \(\ket{\psi(s)}_\sigma\) in the Krylov basis,
\begin{equation}
    \ket{\psi(s)}_\sigma = \sum_n \psi_{n,\sigma}(s) \ket{K_n}_\sigma
\end{equation}
and then define the singular spread complexity,
\begin{equation}
    \mathcal C_\sigma (s) \equiv \sum_n n |\psi_{n,\sigma}(s)|^2
\end{equation}

In the next section, we shall use this singular spread complexity to demarcate chaotic to integrable transition in non-Hermitian systems.

\subsection{B. TFD and CDW Spread Complexity}

In this subsection, we shall be interested in understanding spread complexity dynamics due to the evolution with the actual non-Hermitian Hamiltonian \(\hat{H}\). To calculate spread complexity dynamics of some initial state \(\ket{\psi_0}\), we have to find the suitable Krylov basis vectors and then expand the time-evolved state on that basis. But because of the non-Hermitian nature of the time-evolution generator, the evolution will be non-unitary,
\begin{equation}
    \hat{U}(t) = e^{-i\hat{H}t}\,\,;\,\,\,\, \hat{U}^\dagger (t) \hat{U}(t) \neq 1
\end{equation}
therefore, the time-evolved state has to be obtained by,
\begin{equation}
    \ket{\psi(t)} = \frac{e^{-i\hat{H}t}\ket{\psi_0}}{||e^{-i\hat{H}t}\ket{\psi_0}||}
\end{equation}
where \(||\cdot||\) is the Hilbert space norm.

Since the Hamiltonian is non-Hermitian, we cannot use the Lanczos algorithm to tridiagonalize it. In fact, we will not look forward to tridiagonalizing it, rather we shall use the Arnoldi algorithm to construct the Krylov basis vectors, where the non-Hermitian Hamiltonian will not be a tridiagonal matrix in general. The key point in our purpose is to have an orthonormal Krylov basis \(\{\ket{K_0},\ket{K_1},\ket{K_2},\dots\}\) which have the same span as the non-orthonormal basis \(\{\ket{\psi_0},\hat{H}\ket{\psi_0},\hat{H}^2\ket{\psi_0},\dots\}\). In Arnoldi iteration, we start by defining,
\begin{equation}
    \ket{K_0} \equiv \ket{\psi_0}
\end{equation}
Other Krylov basis vectors for \(n\geq 1\) can be formed by following the steps described below,
\begin{equation}
    \begin{aligned}
        &\ket{w_n} = \hat{H} \ket{K_{n-1}}\\
        &\ket{w'_n} = \ket{w_n} - \sum_{m=0}^{n-1} \bra{K_m} w_n\rangle \ket{K_m}\\
        &\ket{K_n} = \ket{w'_n}/\sqrt{\bra{w'_n}w'_n\rangle}
    \end{aligned}
\end{equation}

Then the time-evolved state can be expanded on this basis,
\begin{equation}
    \ket{\psi(t)} = \sum_{n} \psi_n (t) \ket{K_n}
\end{equation}
with normalized probability of being on the \(n\)th Krylov basis,
\begin{equation}
    p_n(t) = \frac{|\psi_n (t)|^2}{\sum_m |\psi_m (t)|^2}
\end{equation}

Krylov spread complexity is defined by,
\begin{equation}
    \mathcal C_{\mathcal K} (t) = \sum_n n p_n (t)
\end{equation}
this quantity signifies the average position in the Krylov basis.

We also define the Krylov Inverse Participation Ratio (KIPR) by,
\begin{equation}
    \mathcal I_{\mathcal K} (t) = \sum_n p_n (t)^2
\end{equation}
A higher value of KIPR implies localization in the Krylov space.

We apply the above procedure for calculating spread complexity starting from two initial states, which essentially serve different purposes. The first one is analogous to the infinite temperature thermofield double state (TFD),
\begin{equation}
    \ket{\psi_0} = \ket{\psi_{TFD}} = \mathcal N \sum_\alpha \ket{\alpha}_R
\end{equation}
where \(\ket{\alpha}_R\) is the right eigenvector of \(\hat{H}\) with eigenvalue \(E_\alpha\), that is, \(\hat{H}\ket{\alpha}_R = E_\alpha \ket{\alpha}_R\) and \(\mathcal N\) is a suitable normalization.

The other choice of initial state is an initially ordered state which we take to be the charge density wave (CDW) state,
\begin{equation}
    \ket{\psi_0} = \ket{\psi_{CDW}} = \ket{1010\dots1010}
\end{equation}

\section{Results}\label{sec: results}

In this section, we discuss how the different types of spread complexities described in the previous section can be used to understand the non-Hermitian MBL transition in both the models with TRS and without TRS. As it is well-known, such transitions in non-Hermitian systems are sensitive to boundary conditions and we show that the spread complexity dynamics of certain states also capture the differences between the periodic and open boundary conditions. In the going, we also point out the differences between the spread complexity dynamics in non-Hermitian models and spread complexity dynamics in Hermitian models.

\subsection{A. Singular Value Spread Complexity}

\begin{figure}[htbp]
    \centering
    \includegraphics[width=0.9\linewidth]{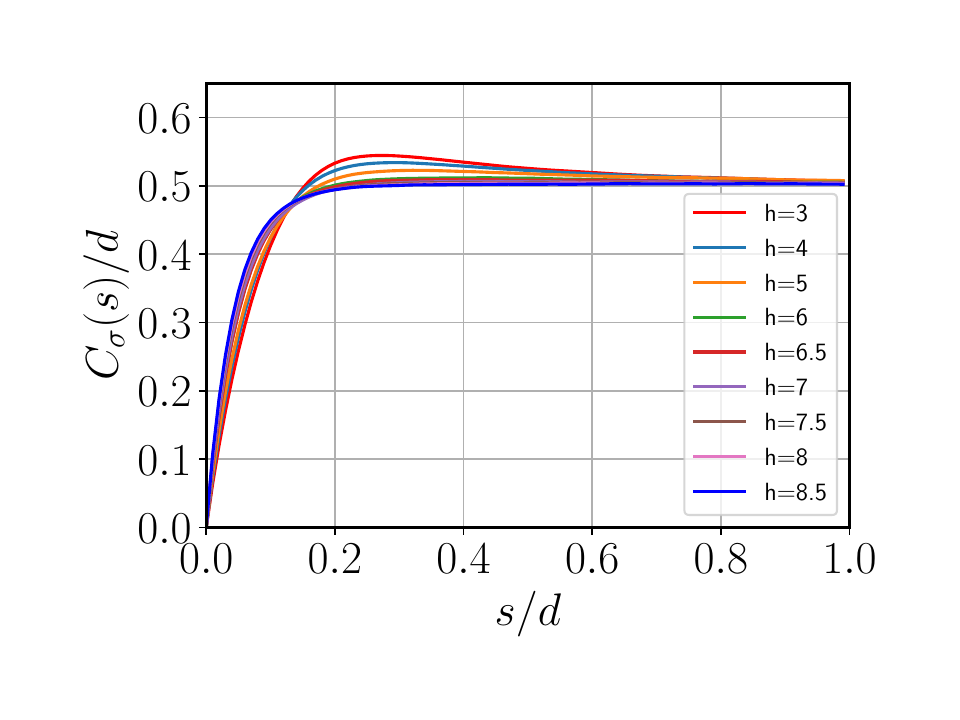}
    \includegraphics[width=0.9\linewidth]{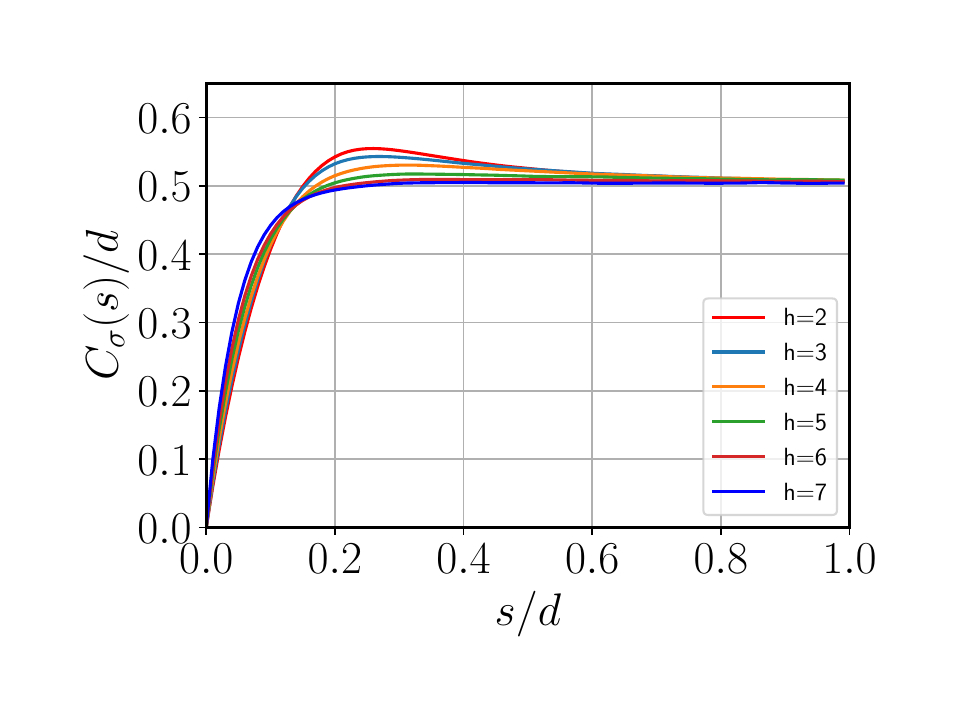}
    \caption{\textbf{Singular value spread complexity in non-Hermitian MBL transition.} \textbf{Top}: singular value spread complexity for the model with TRS, \textbf{Bottom}: singular value spread complexity for the model without TRS. The plots are for \(L=12\) and 250 disorder realizations. Here \(d=924\). We observe that the presence of a peak in the singular value spread complexity can distinguish the chaotic (ergodic) phase from the integrable (here MBL) phase for both models.}
    \label{fig: sing comp}
\end{figure}

We first describe the singular value spread complexity results. We find that the presence of peak in the singular value spread complexity implies a chaotic spectrum (see Fig. \ref{fig: sing comp}). With increasing disorder as the model goes through an MBL transition, the peak gradually diminishes. This observation allows one to use the peak height in the singular value spread complexity as an order parameter for chaotic-to-integrable transitions in non-Hermitian models (for both with and without TRS). However, this transition of peak in singular value spread complexity can be applied to Hermitian models also. Therefore, the notion of singular value spread complexity unifies the non-Hermitian models with Hermitian models for understanding chaotic-to-integrable transition in both kinds of models.

\begin{figure}
    \centering
    \includegraphics[width=0.9\linewidth]{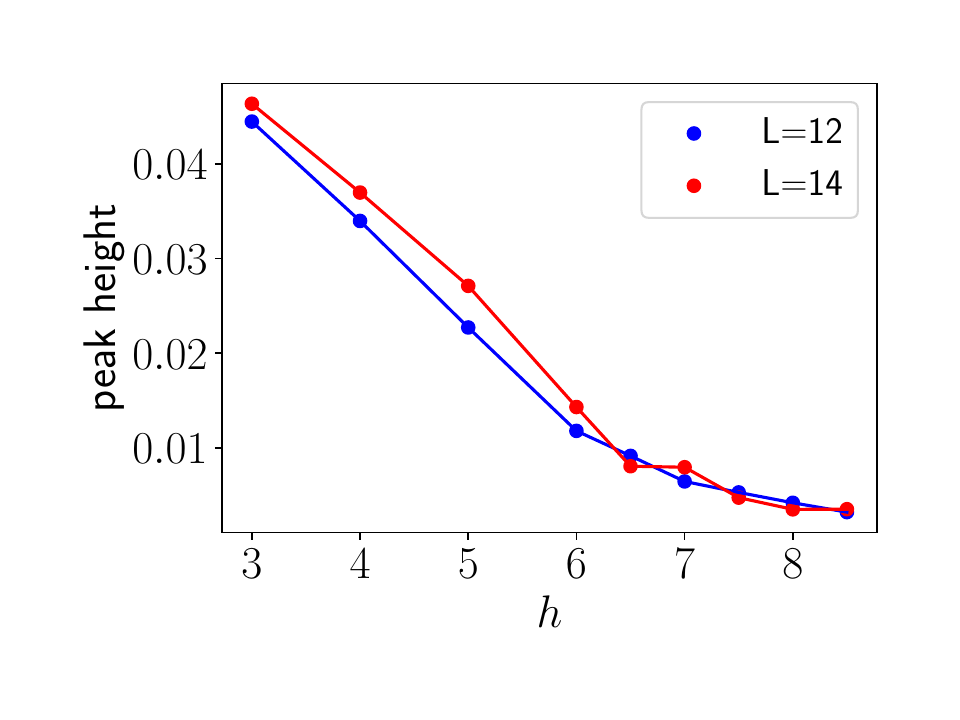}
    \includegraphics[width=0.9\linewidth]{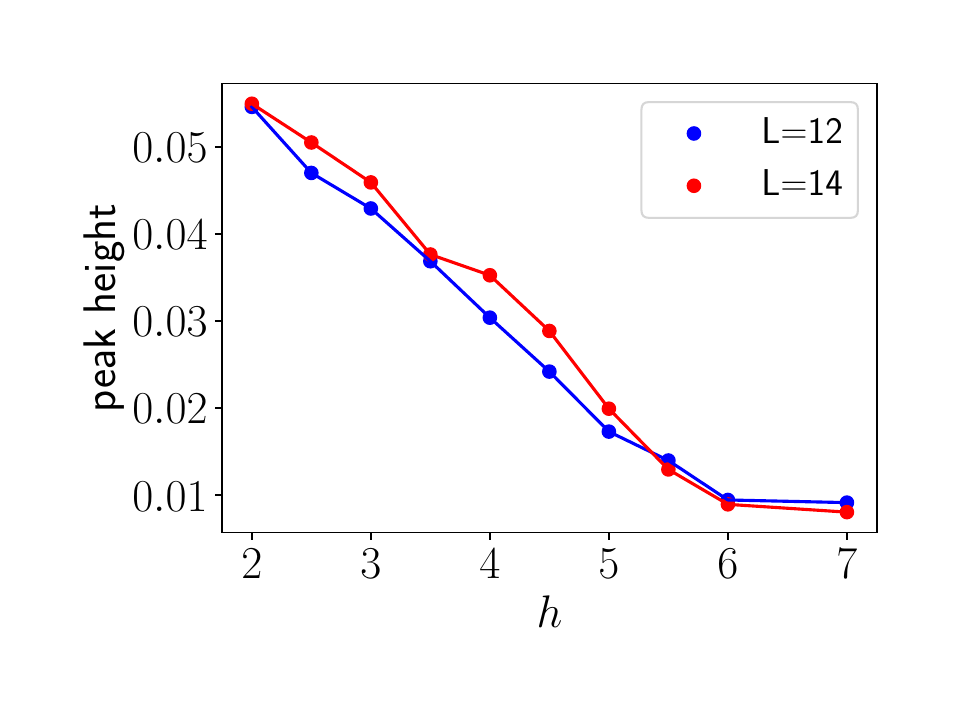}
    \caption{\textbf{Peak height of singular value spread complexity for different disorder strengths} We show the peak height in the singular value spread complexity as a function of disorder strength, \textbf{Top}: for the model with TRS, \textbf{Bottom}: for the model without TRS, for system sizes \(L=12,14\).}
    \label{fig: peak vs h}
\end{figure}

Treating the peak height above 0.5 as an order parameter we can find out the disorder strength at which the chaotic to integrable transitions occur (see Fig.~\ref{fig: peak vs h}).

Another important thing to note from Fig. \ref{fig: sing comp} is the comparison of the peak height of the model with TRS against the model without TRS. The maximum peak height in the model with TRS is less than the maximum peak height in the model without TRS. This is similar to what happens for Hermitian models. In the Hermitian models the peak height in TFD spread complexity of Gaussian Orthogonal Ensemble/GOE (which models the Hermitian chaotic system with TRS), is lesser than the peak height in TFD spread complexity of Gaussian Unitary Ensemble/GUE (which models Hermitian chaotic system without TRS). Therefore we expect this property to hold for peak in singular value spread complexity of non-Hermitian random matrix models, that is for Ginbre Orthogonal Ensemble/GinOE and Ginibre Unitary Ensemble/GinUE. A \(d\) dimensional random matrix belonging to GinOE can be sampled by choosing each element of that matrix randomly drawn from a normal distribution with mean 0 and variance \(1/d\). On the other hand, a \(d\) dimensional random matrix from the GinUE ensemble can be constructed by randomly choosing real and imaginary parts of each element from a normal distribution of mean 0 and variance \(1/2d\).

\begin{figure}
    \centering
    \includegraphics[width=0.9\linewidth]{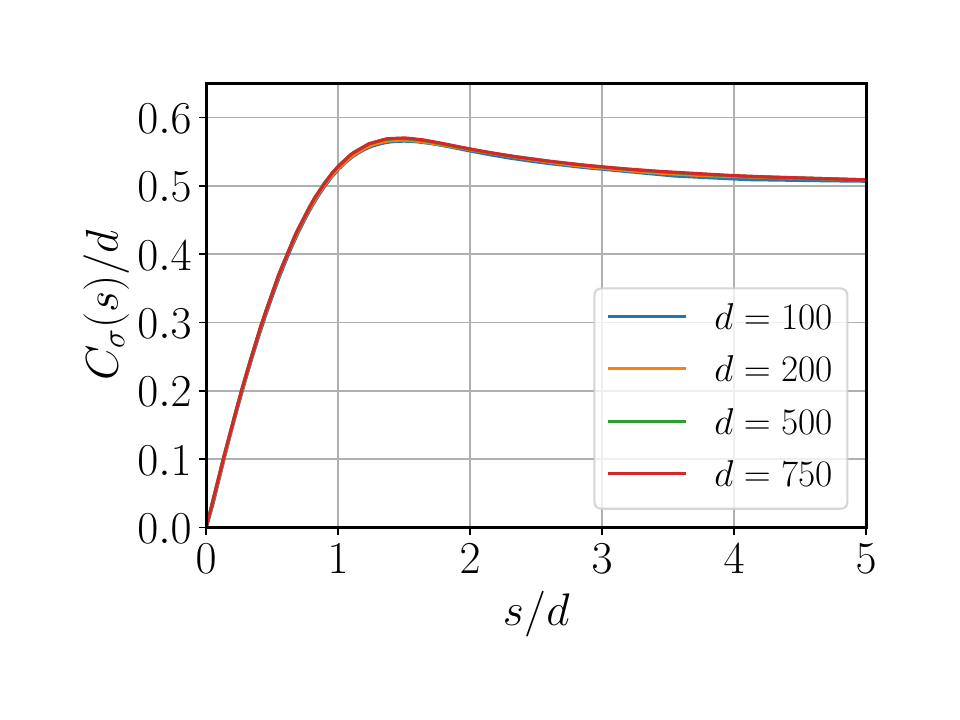}
    \includegraphics[width=0.9\linewidth]{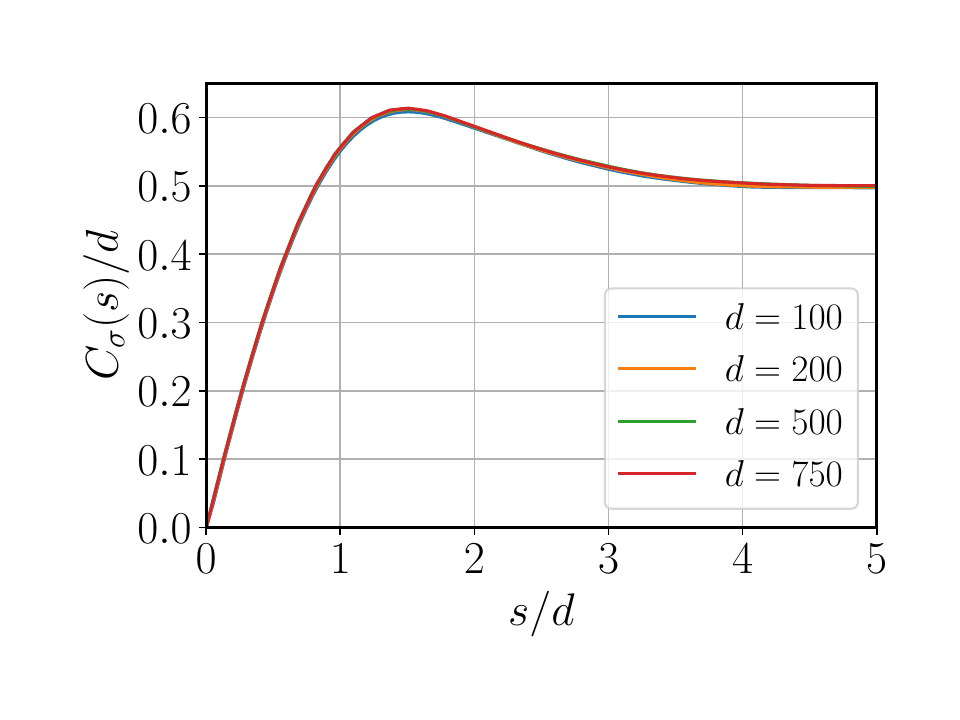}
    \caption{\textbf{Singular value spread complexity in non-Hermitian RMT} Singular value spread complexity for non-Hermitian random matrix theory (RMT),  \textbf{Top}: for GinOE, \textbf{Bottom}: for GinUE for different dimensions \(d\). For each case, \(250\) realizations are taken.7}
    \label{fig: sing comp non herm RMT}
\end{figure}

The singular value spread complexity for both these two ensembles can be calculated by our method. The results are displayed in Fig.~\ref{fig: sing comp non herm RMT}, and we observe that the rescaled complexity curves for different \(d\) collapse to a single curve. Also, the value of peak height is higher in the case of GinUE as compared to GinOE.

\subsection{B. TFD Spread Complexity}

We have seen that the peak in singular value spread complexity can distinguish between the chaotic phase and integrable phase in non-Hermitian models. However, it does not capture the complex-real transition that is present in the model with TRS and is absent in the model without TRS. Interestingly, we find that the spread complexity of infinite temperature TFD state indeed carries the information on how much the fraction of the eigenvalues is complex.

\begin{figure}[htbp]
    \centering
    \includegraphics[width=0.9\linewidth]{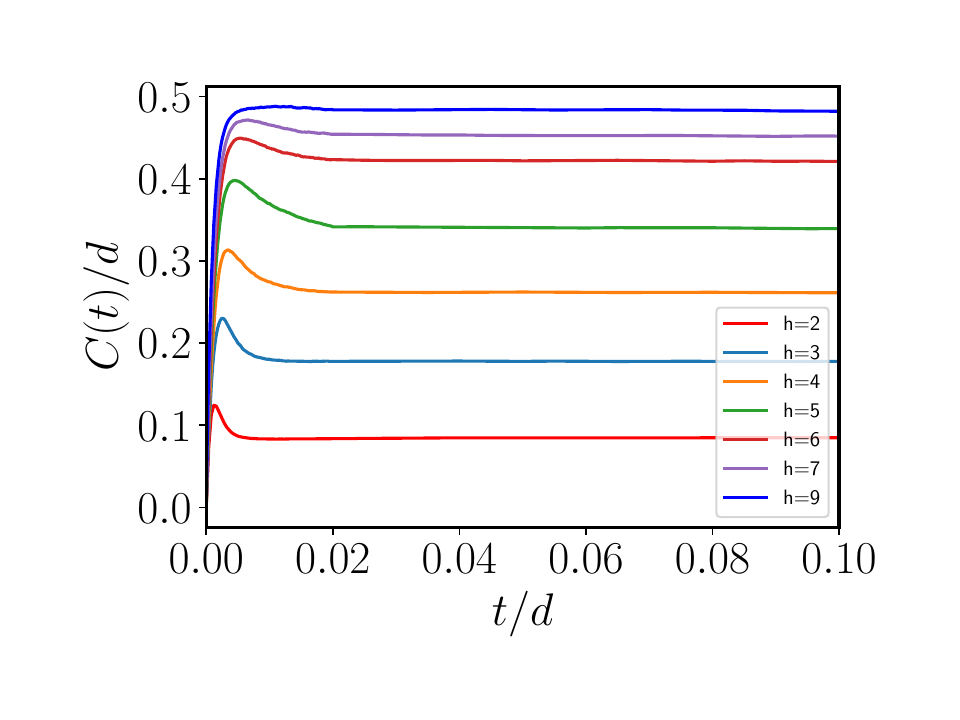}
    \includegraphics[width=0.9\linewidth]{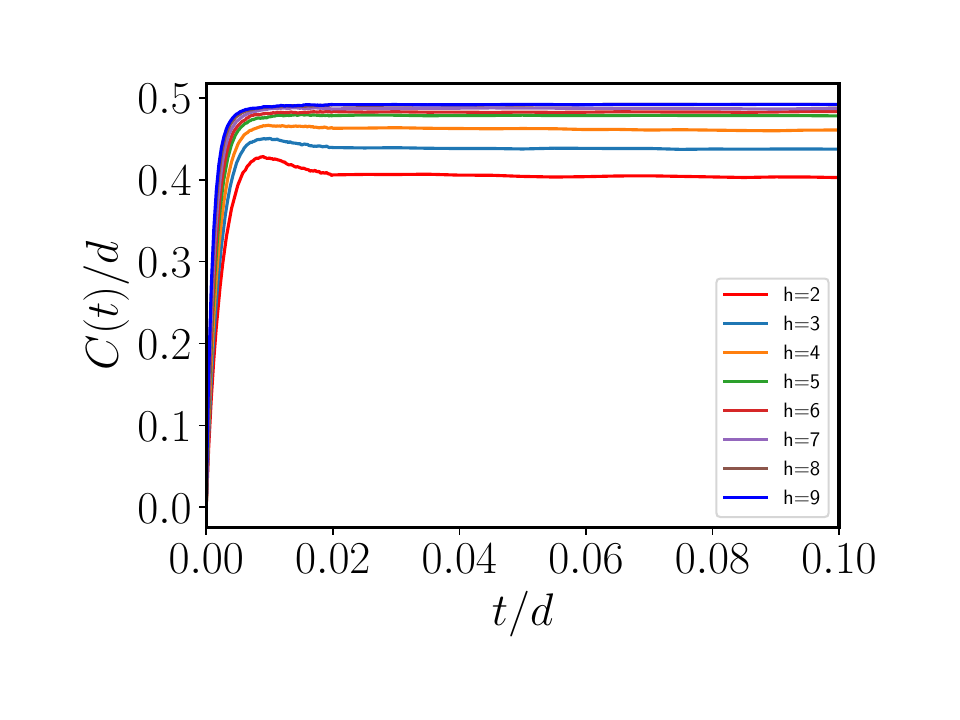}
    \caption{\textbf{TFD spread complexity in the model with TRS.} In the above plots we show the spread complexity dynamics of the infinite temperature TFD state with PBC (top) and OBC (bottom). The saturation value of complexity is directly linked with the proportion of imaginary parts in the eigenvalues of the non-Hermitian Hamiltonian. The eigenvalues are real in the OBC case which explains the higher saturation value compared to PBC at the same disorder strength.
    Since the model with TRS shows complex-real transition, the saturation value increases towards 0.5 by increasing disorder strength. }
    \label{fig: TFD comp with TRS}
\end{figure}

\begin{figure}[htbp]
    \centering
    \includegraphics[width=0.9\linewidth]{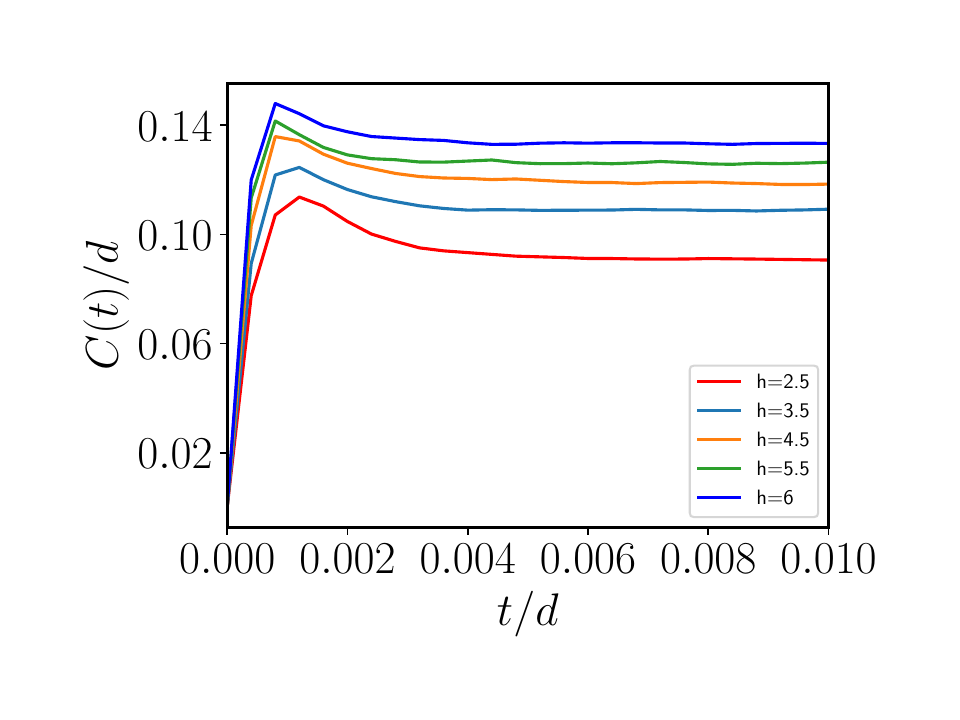}
    \includegraphics[width=0.9\linewidth]{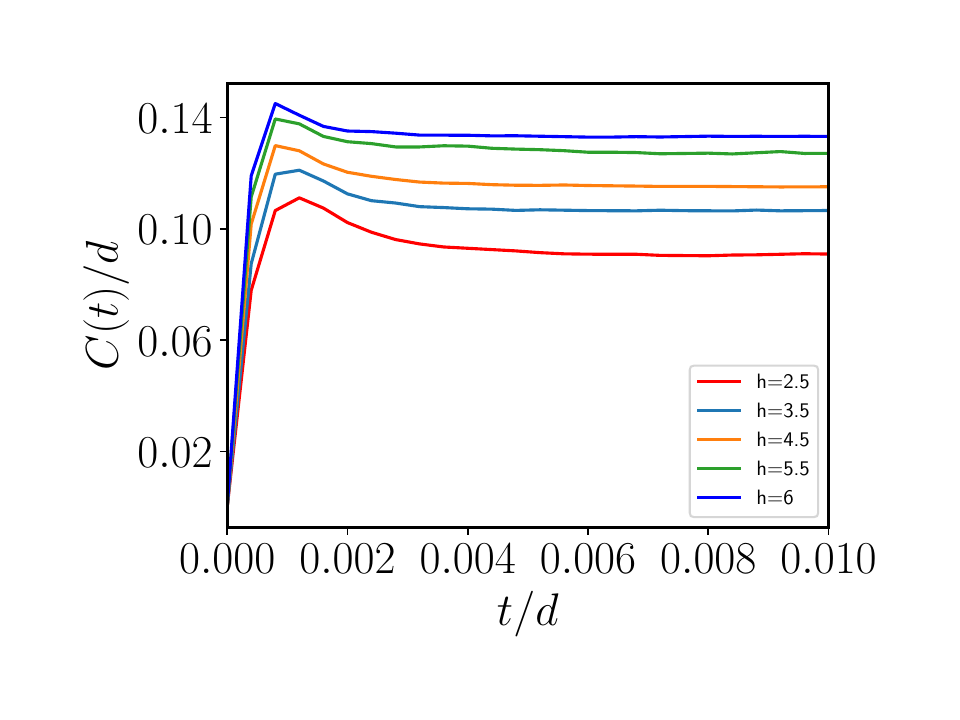}
    \caption{\textbf{TFD spread complexity in the model without TRS.} In the above plots we show the spread complexity of the initial infinite temperature TFD state in the model without TRS for PBC (top) and OBC (bottom). There is no significant difference between the two different boundary conditions. Towards the MBL transition, the peak height decreases and the saturation value increases but saturation never reaches 0.5, because the model without TRS does not show a complex-real transition.}
    \label{fig: TFD comp without TRS}
\end{figure}

Our results show that imaginary parts in the eigenvalue spectra directly affect the saturation value of TFD spread complexity. In particular, the saturation value is always less than 0.5, and the more the fraction of complex eigenvalues, the lesser the saturation value. In the model with TRS (Fig. \ref{fig: TFD comp with TRS}), there is a complex-real transition which explains the observation that the saturation value is going towards 0.5 with increasing disorder. On the other hand, in the model without TRS (Fig. \ref{fig: TFD comp without TRS}) there is no complex-real transition, so the saturation value never approaches 0.5 even with strong disorder.

People have already mentioned that for non-hermitian systems in general boundary conditions play a pivotal role which we also support by our analysis from this spread complexity perspective in these non-hermitian models for both the cases with TRS and without TRS. If we compare the spread complexity for different boundary conditions (PBC and OBC) we observe that these conditions affect the model in the case of TRS more. In fact, the asymmetric hopping model with OBC has no complex eigenvalues than its PBC counterpart which explains the saturation values observed in PBC and OBC cases, in top and bottom plots of Fig. \ref{fig: TFD comp with TRS} respectively. On the other hand, in the particle loss-gain model, boundary conditions do not affect the complexity profile so much, comparing top and bottom plots in Fig. \ref{fig: TFD comp without TRS}.

In this plot chaotic to integrable transition is also reflected in the decreasing height of peak (above the corresponding saturation value) as the system goes through the MBL transition though the transition is not so much prominent. While the peak vanishes completely in the MBL phase of the model with TRS, there is still a small peak even in the MBL phase of the model without TRS.

\subsection{C. CDW Spread Complexity}
Here we describe the results for the spread complexity of an initial charge density wave (CDW) state. This is the analog of N\'eel state complexity considered in a spin model that was considered in Ref.~\cite{Ganguli:2024myj}. There it was found that the spread complexity saturation value of this state can distinguish the ergodic phase from the MBL phase and the reason was the presence of quasi-local integrals of motion (LIOMs) in the MBL phase. Our results here extend this idea to non-Hermitian models with and without TRS. 

\begin{figure}
    \centering
    \includegraphics[width=0.9\linewidth]{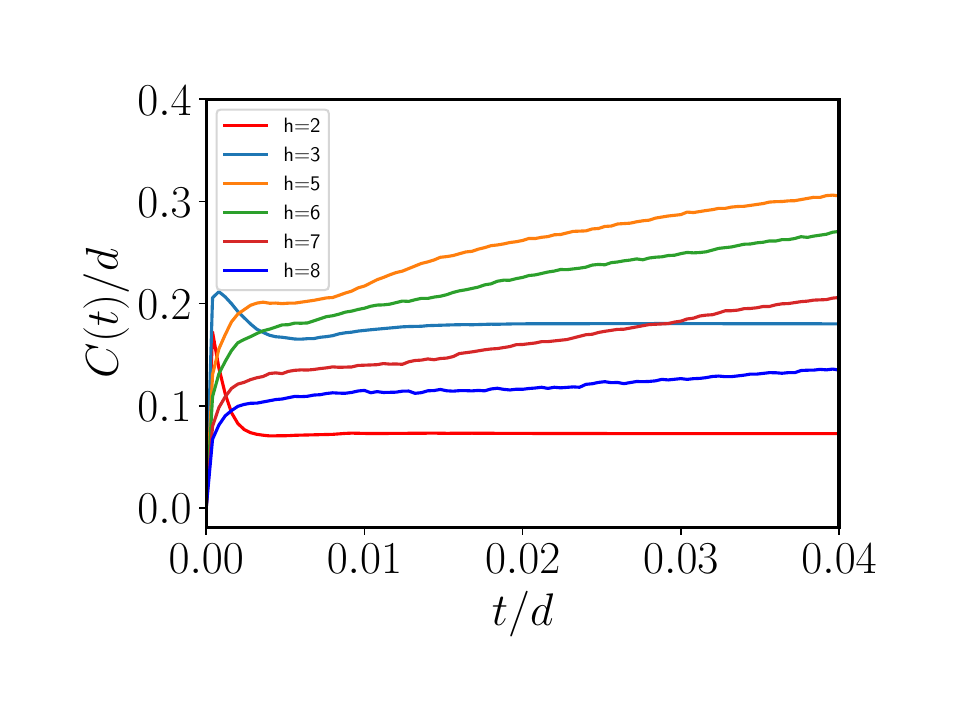}
    \includegraphics[width=0.9\linewidth]{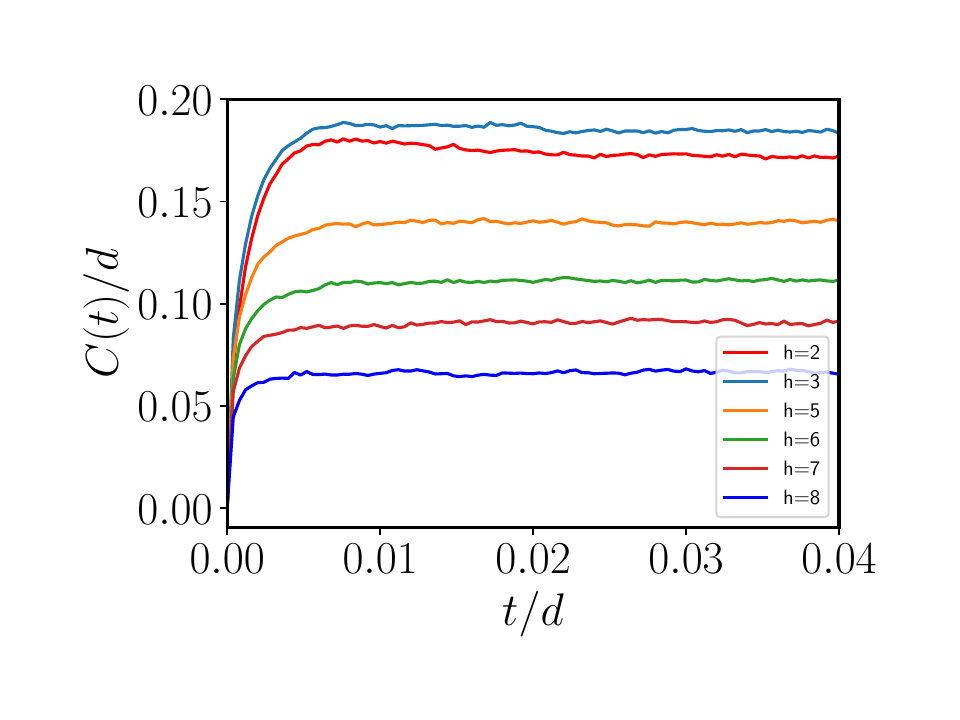}
    \caption{\textbf{CDW spread complexity in the model with TRS.} The above plots show the spread complexity dynamics of the initial charge density wave (CDW) state for the model with TRS for PBC (top) and OBC (bottom). We find that in the MBL phase, there is a suppression complexity which is true for both PBC and OBC even though there is a quantitative difference. In the ergodic phase, we have either a peak in the complexity or a higher saturation value. In the MBL phase, we have lower saturation without any peak.}
    \label{fig: CDW with TRS}
\end{figure}

The plots in Fig. \ref{fig: CDW with TRS} depict the spread complexity of the initial CDW state in the model with TRS with PBC (top plot) and with OBC (bottom plot). We indeed find that in the MBL phase, the saturation value of complexity is less, which is quite evident in the OBC case. In the PBC case, the ergodic phase has a large or small saturation value with a low or high peak respectively, whereas in the MBL phase, the saturation value is small without any peak. Thus CDW spread complexity can distinguish the MBL phase from the ergodic in the model with TRS.

\begin{figure}
    \centering
    \includegraphics[width=0.9\linewidth]{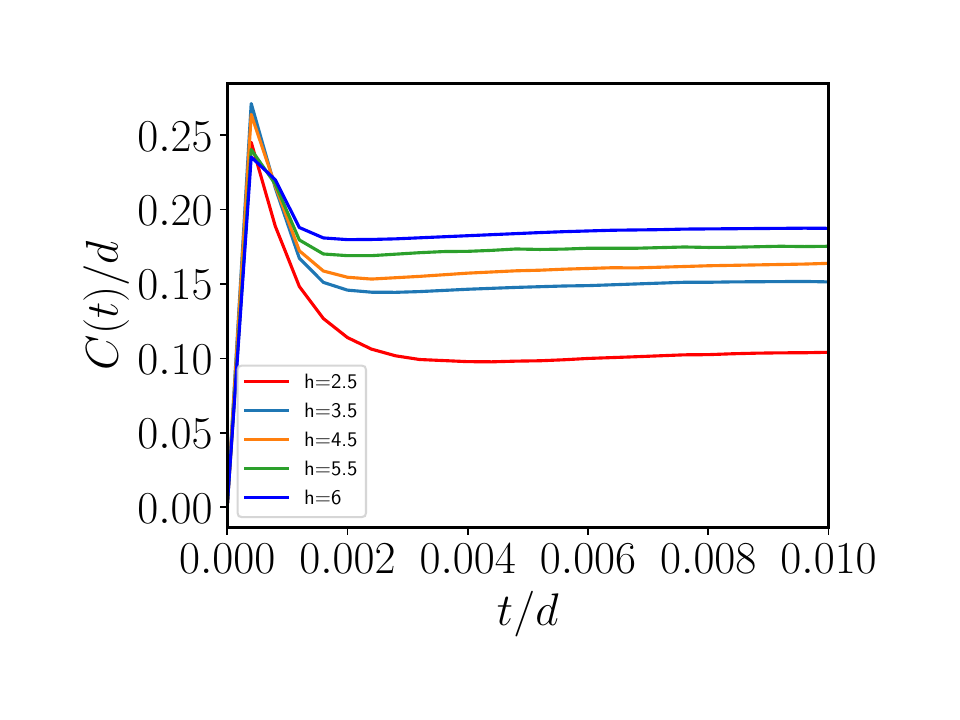}
    \includegraphics[width=0.9\linewidth]{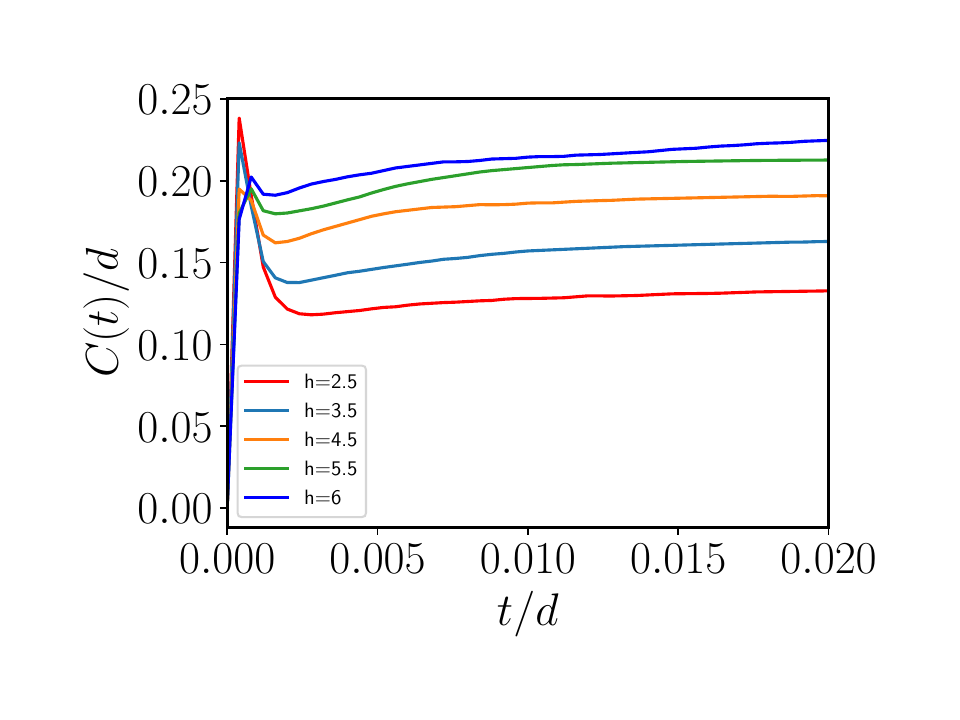}
    \caption{\textbf{CDW spread complexity in the model without TRS.} The above plots show the spread complexity dynamics of the initial charge density wave (CDW) state for the model without TRS for PBC (top) and OBC (bottom). In the model without TRS, the MBL phase has higher saturation with a smaller peak and the ergodic phase has lower saturation with a higher peak for both boundary conditions. However, the distinction is more prominent in the OBC case.}
    \label{fig: CDW without TRS}
\end{figure}

The case for the model without TRS is found to be quite different from the Hermitian model and the non-Hermitian model with TRS. From Fig. \ref{fig: CDW without TRS} we see that the complexity saturation in the MBL phase is \textit{more} than the complexity saturation in the ergodic phase. However, further analytical investigation is needed to explain this behavior. Instead if one considers the peak height (with respect to their corresponding saturation values) then there is a clear transition from high peak value to low peak value as the system goes through the MBL phase transition, for both boundary conditions.

\section{Discussion and Conclusion}\label{sec: discussion}
To summarize, our results show that the singular value spread complexity can be a useful quantity for distinguishing integrable and chaotic phases in non-Hermitian systems and the peak in the singular value spread complexity can act as an order parameter. The presence of peak has also been confirmed in the non-Hermitian random matrix ensembles, GinOE and GinUE. At this point, it is desirable to have a rigorous statistical analysis of singular value spread complexity in the context of integrable to chaotic phase transitions in various other non-Hermitian models. We leave this for future investigations. We have also shown that the saturation value of TFD complexity is sensitive to imaginary parts of the eigenvalues which is evident from the non-hermitian models having TRS and without TRS with different boundary conditions. On the other hand, the spread complexity profile of CDW in the model with TRS can capture the ergodic to MBL phase transition with the MBL phase having lower saturation value. However, the reason behind the higher saturation value of the spread complexity for CDW without TRS in the MBL phase needs further detailed analysis.

\section*{ACKNOWLEDGEMENT} I am grateful to Sumilan Banerjee for useful discussions. I would like to thank Aneek Jana for useful insights, discussions, and collaboration on a related project. This work is supported by the Integrated PhD fellowship of the Indian Institute of Science, Bengaluru.

\paragraph{Note added :}
After completion of this work, \cite{nandy2024krylovspaceapproachsingular} appeared which has a slight overlap with our scope.



\bibliography{bib}

\end{document}